\documentclass[preprint,12pt,showkeys,nofootinbib,floatfix]{revtex4}
\usepackage{amssymb}
\usepackage{amsfonts}
\usepackage{amsmath}
\usepackage{graphicx}
\usepackage{subfigure}
\usepackage[dvipdfm,
            pdfstartview=FitH,
            bookmarksnumbered=true,
            bookmarksopen=true,
            colorlinks,
            pdfborder=001,
            linkcolor=green,
            anchorcolor=green,
            citecolor=red
            ]{hyperref}
\usepackage[toc,page,title,titletoc,header]{appendix}
\usepackage{graphics}
\usepackage{epsfig}
\usepackage{slashed}
\usepackage{latexsym}
\usepackage{syntonly}
\newcommand{\f}{\begin{equation}}
\newcommand{\ff}{\end{equation}}
\newcommand{\fa}{\begin{eqnarray}}
\newcommand{\ffa}{\end{eqnarray}}
\newcounter{fignr}

\begin{document}
\title{Emergent geometry, thermal CFT and surface/state correspondence}
\author{Wen-Cong Gan$^{1,2}$}
\thanks{E-mail address:ganwencong@gmail.com}
\author{Fu-Wen Shu$^{1,2}$}
\thanks{E-mail address:shufuwen@ncu.edu.cn}
\author{Meng-He Wu$^{1,2}$}
\thanks{E-mail address:menghewu.physik@gmail.com}
\affiliation{
$^{1}$Department of Physics, Nanchang University, Nanchang, 330031, China\\
$^{2}$Center for Relativistic Astrophysics and High Energy Physics, Nanchang University, Nanchang 330031, China}
\begin{abstract}

We study a conjectured correspondence between any codimension-two convex surface and a quantum state (SS-duality for short). By generalizing thermofield double formalism to continuum version of the multi-scale entanglement renormalization ansatz (cMERA) and using the SS-duality, we show that thermal geometries naturally emerge as a result of hidden quantum entanglement between two boundary CFTs. We therefore propose a general framework to emerge the thermal geometry from CFT at finite temperature. As an example, the case of $2d$ CFT is considered. We calculate its information metric and show that it is either BTZ black hole or thermal AdS as expected.
\end{abstract}
\maketitle
\section{Introduction}

The fascinating idea that spacetimes might emerge from more fundamental degrees of freedom has attracted more and more attention in the past few years. This idea was revived recently by the discovery of the AdS/CFT correspondence\cite{maldacena1,gkp,witten}. Even though it has lead tremendous progresses in the past few years, fundamental mechanism of the AdS/CFT correspondence still remains a mystery. The situation became better not until the discovery of Ryu-Takayanagi formula\cite{RT}, which states that the entanglement entropy of a subregion $A$ of a $d+1$ dimensional CFT on the boundary of $d+2$ dimensional AdS is proportional to the area of a certain codimension-two extremal surface in the bulk:
\[
S_A=\frac{Area(\gamma_A)}{4G_N^{d+2}}
\]
where $\gamma_A$ is the minimal surface whose boundary coincides the boundary of $A$: $\partial A$.

A recent step for our understanding of holography is made by Miyaji et al in \cite{MT,mnstw}where they proposed a duality called surface/state correspondence (SS-duality). It claims that any codimension two convex surface is dual to a quantum state of a QFT. More precisely, it states that a closed topological trivial convex surface $\Sigma$ is dual to a pure quantum state $|\Phi(\Sigma)\rangle$, while a closed topological non-trivial convex surface $\Sigma$ corresponds to a mixed quantum state $\rho(\Sigma)$. With the help of the SS-duality, one can, in principle, find out the equivalent description of any spacetimes described by Einstein's gravity. In this way, we might encode the information of the boundary QFTs into the bulk geometry, and vice versa.

There is a very different way in mapping states and operators in the boundary Hilbert space to those in the bulk. This is known as the tensor networks. One specific example which is of particular significance is the multi-scale entanglement renormalization ansatz (MERA) \cite{er}. By introducing disentanglers, which remove short-range entanglement, this approach can provide an efficient tensor network description of ground states for critical systems.  The connection between the AdS/CFT and the MERA was first pointed out by Swingle in \cite{swingle}, where he noticed that the renormalization direction along the graph can be viewed as an emergent(discrete) radial direction of the AdS space. This elegant method was latter generalized to continuous version (cMERA), which makes entanglement renormalization available for quantum fields in real space\cite{erqf}. Equipped with this toolkit, the holographic (smooth) geometry can naturally emerge from QFTs\cite{hg}.

Though it is very successful, the full investigation of AdS/cMERA is still very limited. Most past works paid their attention to zero-temperature systems. In this paper, we take a step forward and investigate how to emerge thermal spacetimes from boundary CFT at finite temperature, by making use of cMERA and SS-duality. At first glance the generalization is trivial and one can achieve this as long as the boundary CFT is replaced by a thermal one. However, there are two obstacles that prevent us from this generalization:
First of all, the appearance of black hole (BH) horizon leads to a closed and topologically nontrivial surface in the bulk. This implies, according to the SS-duality, that the dual state in the boundary QFT is no longer a pure state. All calculations must be replaced by thermal mixed states.
Secondly, for finite-temperature CFT, turning on a temperature introduces a scale which screens long-range correlations and the state have thermal correlations in addition to entanglement. One important effect is that the thermal correlations become more relevant as one runs the MERA. The MERA, therefore, truncates at a certain level, which is suggestive of a BH horizon \cite{swingle}.  We often call it the truncated MERA\cite{carrol}. In our previous work\cite{gsw}, we discussed the emergent thermal geometry by generalizing the truncated MERA to continuous one.

An alternative way which is more natural is based on the thermofield double formalism and the emergent tensor network is often called doubled MERA. This proposal \cite{maldacena2}  states that the eternal black hole is dual to two copies of the CFT, in the thermofield double state $|TFD\rangle$. Each asymptotic boundary of AdS is a copy of the original dual CFT. With the help of the SS-duality, we will find by this formulation that the thermal spacetimes naturally appear as a result of hidden quantum entanglement between two boundary CFTs. In this way, we formulate a general framework by which thermal geometries emerge from dual CFTs, without knowing the details of the thermal correlations of the CFT.

\section{Thermofield dynamics and Surface/state correspondence}
\subsection{Thermofield double formalism and doubled cMERA}
We start by introducing a new QFT $H_{tot}$ which is two copies of the original QFT(with Hilbert spaces $H_1$ and $H_2$ respectively). The thermofield double formalism treat the thermal, mixed state $\rho=e^{-\beta H_i}$ ($i=1,2$)
as a pure state in the new double system $H_{tot}=H_1\bigotimes H_2$. Thermofield double state in this doubled system is defined as
\f\label{a2}
|TFD \rangle = \frac{1}{\sqrt{Z(\beta)}} \sum_n e^{-\beta E_n/2}|n\rangle_1 |n\rangle_2,
\ff
where $|n\rangle_1 ,|n\rangle_2$ are energy eigenstates of the two copies of QFT respectively. This is a particular(entangled) pure state in the doubled system. The density matrix of the doubled QFT in this state is
\f\label{rhotot}
\rho_{tot}=|TFD \rangle\langle TFD|.
\ff

 The thermofield double formalism can be applied to the case of the AdS eternal
black hole. The Penrose diagram of an eternal black hole is depicted in Fig.1. The diagram separates the whole spacetime into two asymptotically AdS regions. Each asymptotic boundary of AdS is a copy of the original dual CFT. It is convenient to denote these two identical, non-interacting copies of CFT by CFT$_1$ and CFT$_2$, respectively. According to Maldacena\cite{maldacena2}, this eternal black hole is dual to two copies of the CFT, in the thermofield double state $|TFD\rangle$.
\begin{figure}
  \centering
  \includegraphics[width=.3\textwidth]{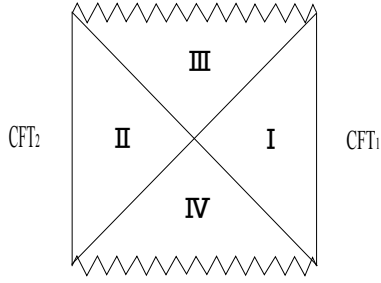}\hspace{1cm}
  \caption{Penrose diagram for an eternal black hole. There are two asymptotically AdS regions which are dual to two copies of CFT.}\label{fig1}
\end{figure}

\begin{figure}
  \centering
  \includegraphics[width=.4\textwidth]{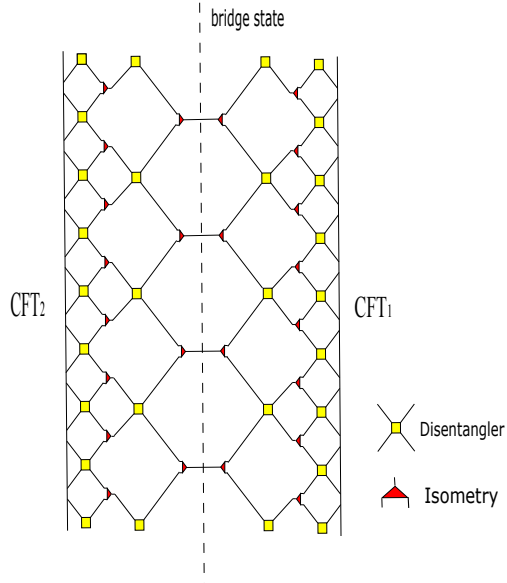}\hspace{1cm}
  \caption{Doubled MERA network. At the center there is a bridge state which glues two copies of the standard MERA. This state is usually viewed as a black hole horizon.}\label{fig2}
\end{figure}

Due to the presence of the horizon, observers in one of those two asymptotically AdS regions (say, region I) cannot come in contact with the other one directly\footnote{However, they connect with each other indirectly through hidden quantum entanglement. The hidden quantum entanglement entropy of the thermal CFT can be viewed as the black hole entropy.}. From the viewpoint of the dual CFTs, for CFT$_1$, information from CFT$_2$ must be traced out. As a consequence, the CFT$_1$ is in a thermal state described by
\f\label{tdm}
\rho_1= \mathrm{Tr}_2 \rho_{tot}= e^{-\beta H_1}.
\ff
The above picture nicely agrees with the MERA at finite temperature as proposed in \cite{hm,jmv1,jmv2,mih}, which is known as the doubled MERA network. It is composed of two copies of the standard MERA for a pure state which are gluing together at infrared points by a "bridge" state. Fig.\ref{fig2} shows a schematic representation of the doubled MERA network. The continuous version of MERA (cMERA) at finite temperature has already been considered in \cite{takayanagi:2014}, where the authors found that, similar to the MERA, finite-temperature cMERA can be constructed by doubling two copies of the standard cMERA.

\subsection{SS-duality description of thermofield dynamics}
Now let us generalize the above picture to a description in terms of the SS-duality. The SS-duality argues a correspondence between any codimension-two convex surface $\Sigma$ and a quantum state of a quantum theory which is dual to the Einstein's gravity. It can be applied to any spacetimes described by Einstein's gravity and therefore can be viewed as a generalization of the AdS/CFT correspondence. More specifically, this duality states that a closed topological trivial convex surface is dual to a pure quantum state $|\Phi(\Sigma)\rangle$, while a closed topological non-trivial convex surface $\Sigma$ corresponds to a mixed quantum state $\rho(\Sigma)$, such as the surface which wraps a black hole. In particular, the zero-size closed surface (i.e. a point) is dual to a boundary state $|B\rangle$\cite{MT,FTE}. When $\Sigma_1$ and $\Sigma_2$ are related by a smooth deformation which preserves convexity, the deformation can be expressed by a unitary transformation
$$
\rho(\Sigma_1)=U(s_1,s_2)\rho(\Sigma_2)U^{-1}(s_1,s_2),
$$
where $U=\mathcal P \exp \{-i\int_{u_2}^{u_1} \hat M(s) \,\mathrm{d}s \}$ with $\mathcal P$ the path-ordering and $M (s)$ a Hermitian operator.

To proceed, let us turn to a cMERA description of the CFT state. From \cite{mnstw}, we learn that a CFT ground state can be expressed in terms of cMERA as follows:
\f\label{vacuum}
|0 \rangle_{CFT} =\mathcal{P} \exp{\left(-i \int_{-\infty}^{0}du \hat K(u)\right)}|I_0 \rangle,
\ff
 where $\hat K(u)$ is the disentangling operator of cMERA at scale $u$, and $|I_0 \rangle$ is the Ishibashi state. The cMERA flow can be adjusted by a conformal transformation. Specifically, for the case of $1+1$ dimensions, it was shown in \cite{mnstw} that one has a transformation $g(\rho,\phi)$ which takes the origin $\rho=0$ to any point ($\rho,\phi$). After acting $g(\rho,\phi)$ transformation, Eq. \eqref{vacuum} can be rewritten as
 \f
|0 \rangle_{CFT} =\mathcal{P} \exp{\left(-i \int_{-\infty}^{0}du \hat K_{(\rho,\phi)}(u)\right)}|I_0 \rangle \equiv U(\rho,\phi)|I_0\rangle,
\ff
where
\f
\hat K_{(\rho,\phi)}(u)=g(\rho,\phi)\hat K(u) g(\rho,\phi)^{-1}.
\ff
Similarly, the CFT excited states $|\Psi_{\alpha}(\rho,\phi)\rangle_{CFT}$ can be expressed in terms of Ishibashi states $|I_{\alpha}\rangle$ for primary field $\Psi_{\alpha}$
\f
|\Psi_{\alpha}(\rho,\phi)\rangle_{CFT}=U_{(\rho,\phi)}|I_{\alpha}\rangle.
\ff
According to surface/state correspondence, there are dualities between quantum states in the CFT and states in the bulk gravity. In particular,
\fa
&&|0\rangle_{CFT}\ \Leftrightarrow \ |0\rangle_{bulk},\\
\label{dualstate}&&|\Psi_{\alpha}(\rho,\phi)\rangle_{CFT}\ \Leftrightarrow \ |\Psi_{\alpha}(\rho,\phi)\rangle_{bulk},
\ffa
 where $|0\rangle_{bulk}\in \mathcal{H}_{bulk}$ is the vacuum state of the bulk gravity, and $|\Psi_{\alpha}(\rho,\phi)\rangle_{bulk}\equiv \hat \psi_\alpha(\rho,\phi)|0 \rangle_{bulk} \in \mathcal{H}_{bulk}$ denotes the locally excited state in the bulk.

 Now we would like to give a SS-duality description of the thermofield dynamics. It is straightforward from Eqs. \eqref{a2} and \eqref{dualstate} to show that
 \fa\label{dualTFD}
&&|TFD \rangle_{CFT} = \frac{1}{\sqrt{Z(\beta)}} \sum_{\alpha} e^{-\beta \Delta_{\alpha}/2}|\Psi_{\alpha}\rangle_{CFT_1} |\Psi_{\alpha}\rangle_{CFT_2} \nonumber\\
\Leftrightarrow \ &&|TFD \rangle_{bulk} = \frac{1}{\sqrt{Z(\beta)}} \sum_{\alpha} e^{-\beta \Delta_{\alpha}/2}|\Psi_{\alpha}\rangle_{bulk_1} |\Psi_{\alpha}\rangle_{bulk_2}
\ffa
where $\Delta_{\alpha}$ is the conformal dimensions of the primary field $\Psi_\alpha$ and we have applied the Maldacena's argument that an eternal black hole can be holographically described by two copies of the CFT in an entangled state. The thermal density matrix in one of the copies of the CFT is obtained by tracing out the contributions of the other copy of the CFT, as shown in \eqref{tdm}. Explicitly, for density matrix of the double CFT given by \eqref{rhotot}, the reduced density matrix of one of the CFT is
\begin{equation}\label{rhoCFT1} \begin{split}
\rho_{CFT_1} &=\mathrm{Tr}_{CFT_2}\ \rho_{tot}      \\
       &=\sum_\gamma {}_{CFT_2}\langle \Psi_\gamma(\rho,\phi)|\Big(\frac{1}{Z(\beta)}\sum_{\alpha ,\alpha'} e^{-\beta \Delta_\alpha /2}|\Psi_\alpha(\rho,\phi)\rangle_{CFT_1}  \\ &\otimes |\Psi_\alpha(\rho,\phi)\rangle_{CFT_2}
       \langle\Psi_{\alpha'}(\rho,\phi)| \otimes {}_{CFT_1}\langle\Psi_{\alpha'}(\rho,\phi)|e^{-\beta \Delta_{\alpha'} /2}\Big) |\Psi_\gamma(\rho,\phi) \rangle_{CFT_2}  \\
       &=\frac{1}{Z(\beta)}\sum_{\alpha,\alpha',\gamma} e^{-\beta \Delta_\alpha /2 -\beta \Delta_{\alpha'} /2}|\Psi_\alpha(\rho,\phi)\rangle_{CFT_1}\langle\Psi_{\alpha'}(\rho,\phi)| \delta_{\gamma\alpha}\delta_{\gamma\alpha'} \\
       &=\frac{1}{Z(\beta)} \sum_{\alpha} e^{-\beta \Delta_\alpha} |\Psi_\alpha(\rho,\phi)\rangle_{CFT_1}\langle\Psi_{\alpha}(\rho,\phi)|,
\end{split} \end{equation}
where $Z(\beta)=\mathrm{Tr} \sum_\alpha e^{-\beta \Delta_\alpha}|\Psi_\alpha(\rho,\phi)\rangle\langle \Psi_\alpha(\rho,\phi)|$.

The above density matrix can be understood in an alternative way. Based on the dualities of the quantum states given by \eqref{dualstate}, we show that thermal states at point $(\rho,\phi)$ in the bulk are dual to thermal states in CFT. Explicitly, the duality is
\fa
\rho_{bulk}=\frac{1}{Z_{bulk}} \sum_\alpha e^{-\beta \Delta_\alpha}|\Psi_\alpha(\rho,\phi)\rangle_{bulk}\langle \Psi_\alpha(\rho,\phi)|\\
\Leftrightarrow\ \rho_{CFT}=\frac{1}{Z}\sum_\alpha e^{-\beta \Delta_\alpha}|\Psi_\alpha(\rho,\phi)\rangle_{CFT}\langle \Psi_\alpha(\rho,\phi)|,\label{rhocft}
\ffa
which is exactly the same as \eqref{rhoCFT1}.

To proceed, let us employ the idea of Fisher information metric
\f\label{db}
ds^2=\mathcal D_B=1-\mathrm{Tr} \sqrt{\rho_1^{1/2} \rho_2 \rho_1^{1/2}},
\ff
where
\fa
\label{rho1}\rho_1&=&\frac{1}{Z} \sum_\alpha e^{-\beta \Delta_\alpha}|\Psi_\alpha (\rho,\phi) \rangle \langle \Psi_\alpha (\rho,\phi)| ,\\
\label{rho2}\rho_2&=&\frac{1}{Z} \sum_{\alpha^{\prime}} e^{-\beta \Delta_{\alpha^{\prime}}}|\Psi_{\alpha^{\prime}} (\rho+d\rho,\phi+d\phi) \rangle \langle \Psi_{\alpha^{\prime}} (\rho+d\rho,\phi+d\phi)|
\ffa
and it measures the distance between two infinitesimally close states $\rho_1$ and $\rho_2$. As usual, we identify this information metric with the metric of the time slice of the emergent spacetime. In this way, the dual geometry of the eternal black hole can be obtained, according to the SS-duality, by considering the distance between two local excitations, which is given by \eqref{db}.

\section{Emergent BTZ black hole}
As an explicit example, in this section we would like to employ our proposal to $2d$ CFT and to see how the expected BTZ black hole is emergent. The quantum distance \eqref{db} plays a significant role in the derivation of the geometry. The key to the calculation of this distance is to find out the form of the states $|\Psi_\alpha (\rho,\phi)\rangle$. For pure AdS, it was shown in \cite{mnstw} that this can be achieved by applying the $SL(2,R)$ isometry and find a set of combinations of the Virasoro generators which preserve the point $\rho=t=0$, then we can obtain the state $|\Psi_\alpha\rangle \equiv |\Psi_\alpha (0,0) \rangle$ and $|\Psi_\alpha (\rho,\phi) \rangle= g(\rho,\phi)|\Psi_\alpha \rangle$, where $g(\rho,\phi)$ is the $SL(2,R)$ transformation which takes the point $\rho=0$ to arbitrary point $(\rho,\phi)$ on $H_2$. With the help of our proposal addressed in the last section, we can generalize this method to CFT at finite temperature.

\begin{figure}
  \centering
  \includegraphics[width=.6\textwidth]{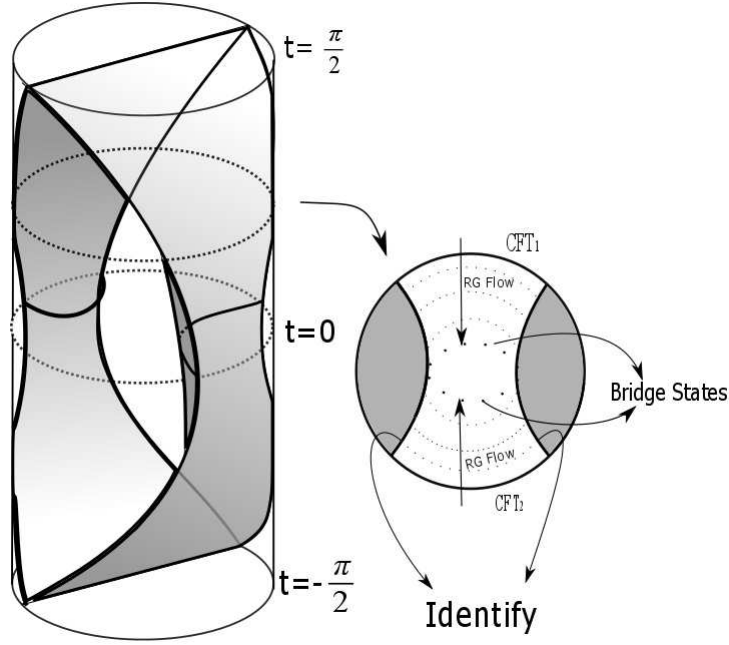}\hspace{0.5cm}
  \caption{The 2+1 dimensional (spinless) BTZ solution in coordinates
    ($t$, $\rho$, $\phi$). All points inside the cylinder belong to anti-de Sitter
    space, its surface ($\rho=1$) representing spatial infinity.
    The BTZ spacetime lies between the two surfaces inside the
    cylinder which are identified under an isometry generated by
    (\ref{L0})-\eqref{Lt}. The RG flow and the horizon (bridge states) are indicated by the dashed lines in
    the constant time slices to the right.}\label{fig3}
\end{figure}

Let us start with a brief review on how to make a BTZ black hole (the AdS black hole in $2+1$ dimensions \cite{btz,bhtz}). Fig. \ref{fig3} shows a sketch of the way in obtaining a BTZ black hole. Roughly speaking, we first find out the ``identification surfaces'' in AdS. These are hypersurfaces that divide the whole spacetime into several regions, some of them have the timelike or null Killing vectors. These regions must be cut out from
anti-de Sitter space to make the identifications permissible. This means that they should
lie entirely within the region where the Killing vector field is space-like. Identifying
corresponding points on these surfaces gives us the BTZ black hole. Before identification, it is a patch of the whole AdS space, whose isometry is given by $SL(2,R)\times SL(2,R)$ generated by the (global) Virasoro generators $(L_1,L_0,L_{-1})$ and $(\tilde L_1,\tilde L_0,\tilde L_{-1})$ of the dual $2d$ CFT. In the global coordinate, they are
\fa\label{L0}
L_0&=&i\partial_+=i\frac{\partial}{\partial x^+},\ \ \tilde L_0=i\partial_-=i\frac{\partial}{\partial x^-},\\
L_{\pm 1}&=&ie^{\pm ix^+}[\frac{\cosh 2\rho}{\sinh 2\rho}\partial_+ -\frac{1}{\sinh 2\rho}\partial_- \mp \frac{i}{2}\partial \rho], \\
\label{Lt}\tilde L_{\pm 1}&=&ie^{\pm ix^-}[\frac{\cosh 2\rho}{\sinh 2\rho}\partial_- -\frac{1}{\sinh 2\rho}\partial_+ \mp \frac{i}{2}\partial \rho],
\ffa
where $x^{\pm}=t\pm\phi$. The identification breaks the symmetry group from $SL(2,R)\times SL(2,R)$ to $SL(2,R)\times U(1)$, however, the BTZ black hole (and its higher dimensional generalization\cite{hdBTZ}) remains locally AdS. Our procedures of deriving the thermal spacetimes in the last section only need local information, it is therefore safe enough to start with \eqref{L0}-\eqref{Lt}.

Following \cite{mnstw}, the excited state $|\Psi_\alpha (\rho,\phi) \rangle$ can be obtained by acting the conformal transformation $g(\rho,\phi)$ to $|\Psi_\alpha\rangle\equiv|\Psi_\alpha (0,0) \rangle$, that is
\f\label{psialpha}
|\Psi_\alpha (\rho,\phi) \rangle= g(\rho,\phi)|\Psi_\alpha \rangle,
\ff
 where $g(\rho,\phi)=e^{i\phi l_0}e^{\frac{\rho}{2}(l_1-l_{-1})}$ with $l_0=L_0- \tilde L_0, l_{-1}=\tilde L_1-L_{-1}, l_1=\tilde L_{-1}-L_1$. The state $|\Psi_\alpha\rangle$ turns out to be of the following form
 \f\label{psialpha1}
 |\Psi_\alpha \rangle \propto e^{i \frac{\pi}{2}(L_0+\tilde L_0)}|J_\alpha \rangle,
 \ff
 where $|J_\alpha \rangle= \sum_{k=0}^{\infty} |k\rangle_L \otimes |k\rangle_R$ are boundary states, and $|k\rangle_L \propto (L_{-1}^k)|\alpha\rangle, |k\rangle_R \propto (\tilde L_{-1}^k)|\alpha\rangle$ are descendants of the primary states $|\alpha\rangle$.

Substituting \eqref{psialpha} and \eqref{psialpha1} into \eqref{rhocft}, making use of \eqref{db}-\eqref{rho2}, we obtains
\begin{align}
&ds^2=\mathcal D_B \nonumber\\
&=1-\sqrt{\frac{1}{Z}\frac{1}{Z(\rho+d\rho,\phi+d\phi)}\sum_{\alpha,\alpha'}e^{-\beta (\Delta_\alpha+\Delta_\alpha')}[1-\frac{1}{8}(d\rho^2+\sinh^2\rho d\phi^2)\langle \Psi_\alpha|l_{-1}l_1+l_1l_{-1}|\Psi_{\alpha'} \rangle]^2} \nonumber\\
&=1-f(\beta)[1-\frac{1}{8\delta^2}(d\rho^2+\sinh^2\rho d\phi^2)],
\end{align}
where the following relations have been employed
\begin{align}
\langle \Psi_\alpha|l_{-1}l_1+l_1l_{-1}|\Psi_{\alpha'} \rangle \propto \langle \Psi_\alpha|l_{-1}l_1+l_1l_{-1}|\Psi_{\alpha}\rangle\delta_{\alpha\alpha'},\nonumber
\end{align}
and
\begin{align}
\langle \Psi_\alpha|l_{-1}l_1+l_1l_{-1}|\Psi_{\alpha} \rangle \simeq \frac{1}{\delta^2}. \nonumber
\end{align}
Noticing that
\begin{align}
f(\beta)=\frac{1}{Z}\frac{1}{Z(\rho+d\rho,\phi+d\phi)}\sum_{\alpha,\alpha'}e^{-\beta (\Delta_\alpha+\Delta_\alpha')} \simeq 1,\nonumber
\end{align}
we get the metric of time slice
\f\label{thermalAds}
ds^2\simeq \frac{f(\beta)}{8\delta^2}(d\rho^2+\sinh^2\rho d\phi^2).
\ff
This is exactly the spatial part of the AdS metric in global coordinate, but now it involves temperature through the parameter $\beta$, and can be viewed as thermal AdS.  If we define a set of new coordinates
$$
r=r_+\cosh \rho,\ \hat{t}=\frac{i\sqrt{f(\beta)}\phi}{2\sqrt{2}r_+ \delta},\ \theta=\frac{\sqrt{f(\beta)}t}{2\sqrt{2}r_+\delta},
$$
the full thermal metric \eqref{thermalAds} after adding $g_{tt}$ can be recast as
\f\label{btz}
ds^2=-\left(r^2-r_+^2\right)d\hat{t}^2 +\frac{f(\beta)}{8\delta^2} \frac{dr^2}{r^2-r_+^2} + r^2 d\theta^2.
\ff
This is exactly the BTZ metric as expected. With the help of the thermofield double formalism, above procedures allow us knowing little information about the thermal CFT, which is one of the main merits of the proposal.

\section{Conclusions}
In this paper we have studied emergent geometries from CFT at finite temperature in the setup of the surface/state correspondence. We propose a general framework through which thermal geometries emerge from boundary CFTs.  Instead of introducing a truncated level to the MERA tensor network, our proposal is realized by applying the thermofield double formalism to the SS-duality, and the thermal correlations are read off by tracing over one of the copies of the CFT. The main advantage of this framework is that the details of the thermal correlations of the CFT are not required.  As an explicit example, we computed the information metric for a locally excited mixed state of two dimensional CFT at finite temperature, and showed that the emergent spacetimes are either thermal AdS or BTZ black hole as expected.

In the present framework, the information metric only depends on the behavior of two nearby states. This implies, according to the SS-duality, emergent bulk metric can be obtained by merely knowing the local information. This is one of the advantages of this framework. However, the other side of the coin is that nonlocal information is missing and the global symmetry cannot emerge from local operators. Although it is important, it is a tough difficulty and is out of the scope of the present paper. One possible clue is to resort to the kinematic space as shown in \cite{czech1,czech2}, where the authors developed tools for constructing local bulk operators in terms of non-local objects in the CFT.

Another important future problem which is of close correlation is to find which factor determines the emergent spacetime to be thermal AdS or BTZ black hole, or equivalently, the Hawking-Page phase transition. In our previous work\cite{gsw} we have found a cMERA description of the Hawking-Page phase transition in the framework of the truncated MERA. In the present framework, however, its solution obviously depends on the details of the global behavior, which is far from being achieved currently. We hope in the future work we can find its description in this framework.

\begin{acknowledgments}
We are very grateful to Bartlomiej Czech and Tadashi Takayanagi for careful reading of the first version of this paper and giving us valuable comments.
We also thank Xi Dong, Kanato Goto, Yuting Hu, Yi Ling, Masamichi Miyaji, Xiao-Liang Qi, Edward Witten and Shao-Feng Wu for useful conversations. We are grateful for stimulating discussions to participants of ``Holographic duality for condensed matter physics'' held in KITPC at CAS and the conference ``Strings 2016'' held in YMSC, Tsinghua. This work was supported in part by the National Natural Science Foundation of China under Grant No. 11465012, the Natural Science Foundation of Jiangxi Province under Grant No. 20142BAB202007 and the 555 talent project of Jiangxi Province.
\end{acknowledgments}

\end{document}